\documentclass[11pt]{article}
\usepackage{epsfig}
\newcommand{\be}{\begin{equation}}
\newcommand{\ee}{\end{equation}}
\newcommand{\ba}{\begin{eqnarray}}
\newcommand{\ea}{\end{eqnarray}}
\newcommand{\bea}{\begin{eqnarray}}
\newcommand{\eea}{\end{eqnarray}}
\begin{document}
%\begin{flushright}
%Hyperon-06 : \today 
%\end{flushright} 

\begin{center} 

{\Large\bf Comments on the electromagnetic 
decay of $\Lambda (1520)$
%excited hyperons 
}

\vskip 5mm 

\large{
F. Myhrer\footnote{E-mail:myhrer@physics.sc.edu} 
}

\vskip 5mm 

{\it 
%${}^{(a)}$ 
Department of Physics and Astronomy, \\ 
University of South Carolina,
Columbia, 
SC 29208, USA 
}
\end{center} 

\vskip 5mm 

%%%%%%%%%%%%%%%%%%%%%%%%%%%%%%%%%%%%%%%%%%%%%%%%%%%%%%%%%%%% 

Abstract: \\ 

The electromagnetic decay processes of 
excited hyperon states are 
a very sensitive probe of the structure 
of hyperons. 
We will argue that the recent measurements of   
electromagnetic decay rates  
 indicate that the 
wave functions of 
the hyperon ground states should 
contain sizeable components of excited quark states 
(configuration mixing). 
Flavor-$SU(3)$ is a broken symmetry and  
it appears that the  
hyperon wave functions should preferably be 
written in a  
$uds$-basis, where only the light $u$ and $d$ quarks are symmetrized, 
and not in the usual $SU(6)$-basis.

%%%%%%%%%%%%%%%%%%%%%%%%%%%%%%%%%%%%%%%%%%%%%%%%%%%%%%%%%%%%%%%%

\newpage 

\vspace{10mm} 

\renewcommand{\thefootnote}{\arabic{footnote}} 
\setcounter{footnote}{0} 

%%%%%%%%%%%%%%%%%%%%%%%%%%
%\section
{\bf Introduction}
%%%%%%%%%%%%%%%%%%%%%%%%%% 

\vspace{3mm} 

The electromagnetic transition rates of excited baryons 
to their respective ground states offer a 
stringent test on the quark model dynamics, 
since electromagnetic transitions 
provide a relatively clean 
probe of the wave functions of the 
initial and final baryon states. 
It is therefore 
highly desirable to measure the electromagnetic 
decay rates from excited baryon states in order to 
refine the quark model description of the baryons.  
% 
%%%%%%%%%%%%%%%%%%%%%%%%%%%%%%%%%%%%%%%%%%%%%%%%%%%%%% 
% 
To date very few electromagnetic 
transition rates have been 
measured for the excited baryon resonances. 
For a detailed discussion of the experimental and 
theoretical status of the excited baryons 
and their electromagnetic decays, 
see the review by Landsberg \cite{landsberg96}. 
The radiative decays of excited hyperon states have 
very small branching ratios, and   
the experimental determination of  
the $\Lambda (1520)$ radiative 
decay rate  
%$\Gamma_\gamma$ 
is quite an accomplishment.  
One therefore naturally expects these difficult 
measurements of the radiative transition rates 
to have 
relatively large 
experimental uncertainties. 
Fortunately, despite the large uncertainties, these rates 
can be very decisive in discriminating among  
the different quark model predictions.   
% 
%%%%%%%%%%%%%%%%%%%%%%%%%%%%%%%%%%%%%%%%%%%%%%%%%%%%% 
% 
Recently the CLAS collaboration 
at Jefferson Lab 
published a new measurement of the radiative 
decay widths of the excited hyperons 
$\Sigma^0 (1385)$ and 
$\Lambda (1520)$ \cite{CLAS05}. 
The measured decay rate for 
$\Lambda (1520) \to \gamma \; \Lambda (1116) $, 
$\Gamma_\gamma = 167 \pm 43 \; ^{+26}_{-12}$ keV~\cite{CLAS05}, 
confirms the recent result of 
Antipov {\it et al.} \cite{russian04} 
and the rate measured a long time ago by 
Mast {\it et al.} \cite{mast68}. 
These three experiments establish the decay rate to be 
$\Gamma_\gamma > 100$keV, 
which is in contradiction 
with $\Gamma_\gamma < 50$keV found by  
Bertini {\it et al.} \cite{bertini84, bertini87}. 
%
%%%%%%%%%%%%%%%%%%%%%%%%%%%%%%%%%%%%%%%%%%%%%%%%%%%%% 
%
As will be argued, despite the 
large errors,  
the measured rate gives 
valuable clues regarding 
the structure of the hyperon wave functions. 

\vspace{3mm} 

In this short note we concentrate the discussion on 
the measured 
$\Lambda (1520) \to \gamma \; \Lambda (1116) $  
decay rate $\Gamma_\gamma$.  
In the theoretical evaluation of the electromagnetic  
transition rates the usual assumption 
is that the excited hyperon states are 
three-quark states. 
It has been argued by many that  
the lowest lying 
excited hyperon states, e.g. $\Lambda (1405)$, 
could contain a dominant 
$\bar{K} N$ component (a ``$\bar{q} q^4$"-admixture) 
due to the nearby 
$\bar{K} N$ threshold, 
see e.g. \cite{Dalitz}. 
%,others,Weise}. 
A $\bar{q} q^4$-admixture 
in the excited hyperon wave function 
would modify the predicted 
electromagnetic transition rates. 
In the discussion below we will not consider  
this complication but 
we will comment on this aspect at the end of this paper.  
We will assume  
that 
$\Lambda (1520)$ is predominantly a 
three-quark state.  
The experimental value of 
$\Gamma_\gamma$~\cite{CLAS05,russian04,mast68}  
is 
consistent with the predictions of a 
particular 
non-relativistic quark model (NRQM) 
and a relativized version of the NRQM 
\cite{dhk83,wpr91,kms85}. 
The measured rate differs markedly from, for example, 
the various bag-model 
evaluations of $\Gamma_\gamma$ \cite{kms85,um91,um93}. 
% 
%%%%%%%%%%%%%%%%%%%%%%%%%%%%%%%%%%%%%%%%%%%%%%%%%%%%%%%%
% 
As will be argued, there are two 
principal reasons why 
all bag models fail to explain the large observed 
$\Gamma_\gamma$ value. 
Below we 
will also discuss the reason why some potential model 
evaluations of $\Gamma_\gamma$ 
give too small a value for this rate, 
and 
comment on  some recent model calculations of 
$\Gamma_\gamma$~\cite{bonn06,oset06,Yu06}.  
We will show that the 
experiments \cite{CLAS05,russian04,mast68} 
%despite the large experimental uncertainty  
%e.g. Ref.\cite{CLAS05} finds 
%$\Gamma_\gamma \simeq 167 \pm 43 \; ^{+26}_{-12}$ keV, 
give 
a clear indication 
that the ground state $\Lambda (1116)$ contains 
configuration mixing from excited quark-states. 

\vspace{3mm}
%
%\newpage

%%%%%%%%%%%%%%%%%%%%%%%%%%%%%%%%%%%%%%%%%%%%%%%%%%%%%%
%\vspace{3mm} \noindent 
%\section
{\bf The hyperon wave functions } 
%%%%%%%%%%%%%%%%%%%%%%%%%%%%%%%%%%%%%%%%%%%%%%%%%%%%%%

\vspace{3mm}

In order to discuss the electromagnetic transition 
$\Lambda (1520) \to \gamma \; \Lambda (1116) $,   
we first present the wave functions of these 
two hyperon states. %which are used in various models. 
%
%%%%%%%%%%%%%%%%%%%%%%%%%%%%%%%%%%%%%%%%%%%%%%%%%%%%%%  
%
The ground state, 
$\Lambda (1116)$, is often assumed to be 
the simple 
$SU(6)$ spin-flavor spatially-symmetric  
state, 
%\footnote{ as used by, e.g. \cite{dhk83}:} 
\bea
|\Lambda(1116)\rangle 
 &=& 
|{\bf 8}, ^2S_S\rangle \; , 
\label{eq:Lsymm}
\eea 
i.e. the three 
quarks are all in the lowest 
energy $s$-state  
forming a spatially symmetric ($S$),  
flavor $SU(3)$-octet (${\bf 8}$), spin doublet state. 

\vspace{3mm}

The NRQM of 
Isgur and Karl \cite{ik77,ik78} generated 
a $\Lambda (1116)$ 
three-quark ground state 
wave function in the $uds$-basis 
instead of the common $SU(6)$-basis. 
The rationale is that the heavy $s$-quark 
mass breaks flavor 
$SU(3)$ symmetry ($m_u=m_d \ne m_s$).  
In the $uds$-basis only the 
equal mass $u$ and $d$ 
quarks are symmetrized.  
Isgur and Karl assumed a spin-independent 
harmonic oscillator (HO) confining potential, 
and the two   
HO frequencies associated with the two 
Jacobi coordinates, $\rho$ and $\lambda$, 
are different since $m_s > m_d = m_u$, 
see also Ref.\cite{orsay78}. 
The spatial wave function for the $s$-quark, 
which is associated with the $\lambda$ coordinate,
is therefore not the same as the 
$u - d$ wave function characterized by  
the $\rho$ coordinate. 
Furthermore, 
Isgur and Karl showed that in the NRQM the 
effective color hyperfine interaction, where they 
included only a local spin-spin and a 
tensor interaction among pairs of quarks,  
couples the quark ground state to 
higher energy HO-states,  
$S^\prime$ and $D$, and modifies the 
$\Lambda (1116)$ state, Eq.(\ref{eq:Lsymm}). 
This generates a 
more refined description of the 
$\Lambda (1116)$ state wave function and 
these extra wave function components  
give important contributions to the value 
of $\Gamma_\gamma$.  
The modified  $\Lambda (1116)$ wave function is 
(see e.g. Ref.~\cite{ki80}): 
\bea
|\Lambda(1116)\rangle^\prime  
 &=& 
0.93 |{\bf 8}, ^2S_S\rangle 
- 0.30 |{\bf 8}, ^2S^\prime_S\rangle 
- 0.20 |{\bf 8}, ^2S_M\rangle 
- 0.05 |{\bf 1}, ^2S_M\rangle 
- 0.03 |{\bf 8}, ^4D_M\rangle \; . 
\label{eq:Lmixed}
\eea 
Here ${\bf 8} $ denotes a 
flavor-octet state,
the notation $^{(2S+1)}L_\sigma$ specifies the  
spin-multiplet $2S+1$ and the total 
orbital angular momentum $L$ in a  
spatial symmetry-state 
$\sigma$ with total 
angular momentum and parity 
$J^P= |\vec{L}+\vec{S}|^P=1/2^+$.  
The symmetries of the spatial wave function 
is either a symmetric, 
$\sigma=S$, or a 
mixed symmetry wave function, $\sigma =M$. 
Thus, according to NRQM of Isgur and Karl,  
$|\Lambda(1116)\rangle^\prime $ is 
predominantly a flavor-octet, {\bf 8}, spatially 
symmetric state with smaller admixtures from 
the excited HO  
($L=0$) $S^\prime$ and ($L=2$) $D$-states. 
The configuration admixtures in Eq.(\ref{eq:Lmixed}) 
affect strongly the calculated rate $\Gamma_\gamma$ 
as we will discuss.  

\vspace{3mm} 

%%%%%%%%%%%%%%%%%%%%%%%%%%%%%%%%%%%%%%%%%%%%%
%
The first excited 
hyperon three-quark states, $\Lambda^*$, have one 
quark in a $p$-state 
%, $l=1$, 
%(with angular momentum  $j=1/2$ or $j=3/2$) 
and the two others in 
the lowest-energy $s$-state. 
% i.e. $j=1/2$. 
% 
When the spatial quark wave functions 
are Gaussians, e.g. Ref.~\cite{ik78}, 
the separation of the 
center-of-mass coordinate (c.o.m.)  
and the baryon's internal Jacobi coordinates 
appears trivially. 
In the HO model of Ref.~\cite{ik78}  this  
internal wave function of the  
three-quark excited hyperon 
describes a pair of 
quarks in relative $s$-state with    
the third quark ``orbiting" the two 
in a $p$-state. 
As explicitly shown by Isgur and Karl \cite{ik78}, 
the usual $SU(6)$-states for excited baryons are 
only recovered in the limit of equal quark masses, i.e. 
$m_s\to m_d=m_u$.\footnote{
This means both Eq.(\ref{eq:Lsymm}) and 
Eq.(\ref{eq:Lmixed}) are approximations of 
the {\it uds} basis, see e.g., 
section III in the second paper in 
Ref.~\cite{ik78}. 
This reference also says that most (excited) 
physical states are closer to pure 
$\rho$ or $\lambda$ states than to pure 
$SU(3)$ eigenstates. 
}
The usual assumption 
is that the negative parity 
$\Lambda^* $ states 
belong to the {\bf 70}-plet of the 
spin-flavor $SU(6)$ symmetry where the  
wave functions of the 
$\Lambda^*$ states 
have mixed spatial symmetry and 
contain 
substantial $SU(3)$-multiplet mixing. 
In the SU(6)-basis, the three-quark 
$\Lambda^*$ states 
combine  the flavor-singlet ${\bf 1}$ state 
with a spin-doublet, $2S+1=2$, state whereas  
the  flavor octet ${\bf 8}$ 
can be in either a 
spin doublet or a spin-quartet, $2S+1=4$, state.  
This assumption leads to the following 
generic wave function for the 
$\Lambda (1520)$ $J^P=3/2^-$ state: 
\bea
|\Lambda^*(J^P) \rangle &=& a\; |^2{\bf 1}_J \rangle + 
b\; |^4{\bf 8}_J\rangle +c\; |^2{\bf 8}_J\rangle \; .  
\label{eq:L1520} 
\eea
As discussed below, 
the different quark models result in different 
values for the coefficients $a, b,$ and $c$. 
%
%\vspace{3mm} 
%
A few comments regarding the exclusion of 
the $SU(6)$ {\bf 56}-plet 
are necessary for the discussion below. 
Historically, the 
spatially symmetric 
wave function of 
two quarks in $s$-states and one in a 
$p$-state is 
considered to be the 
translational mode of the simple 
baryon ground state,  e.g. 
$\Lambda (1116)$  Eq.(\ref{eq:Lsymm}),  
where all three quarks 
are in the lowest $s$-state, see e.g. 
Ref.~\cite{close79}.   
The spatially symmetric 
three-quark wave function belongs 
to the $SU(6)$ {\bf 56}-plet, and 
one therefore assumes that 
the negative parity 
$\Lambda^*$ states 
are described by the spatially mixed-symmetry 
three-quark wave functions of the 
{\bf 70}-plet. 
The assumption that the 
excited negative-parity baryons 
states belong to the $SU(6)$ {\bf 70}-plet  
is adopted  
in many quark models. 
However, the argument that the {\bf 56}-plet 
describes the translation of the ground state  
has to be reconsidered 
if the ground state 
baryons contain other spatial configurations, see    
e.g. Eq.(\ref{eq:Lmixed}).  
Using the MIT-bag model Rebbi \cite{Rebbi75}   
investigated the argument that 
the {\bf 56}-plet could describe 
the c.o.m. motion of a baryon 
and, despite   
the fact that in 
the static cavity approximation the bag is 
not a momentum eigenstate, Rebbi   
found that this argument is reasonable.   
(This argument was re-examined 
in the cloudy bag \cite{um89} 
and found to be questionable.)

\vspace{3mm} 

%%%%%%%%%%%%%%%%%%%%%%%%%%%%%%%%%%%%%%%%%%%%%%%%%%%%
%
{\bf %non-relativistic 
Quark models} 
%%%%%%%%%%%%%%%%%%%%%%%%%%%%%%%%%%%%%%%%%%%%%%%%%%% 

\vspace{3mm} 

The quark wave-functions of the excited baryon states 
are determined  
using a {\it spin-independent} 
harmonic oscillator (HO) potential as the confining 
potential in the NRQM. 
The NRQM also includes a spin-dependent potential  
which is 
derived from the 
one-gluon-exchange (OGE) between quarks. 
This OGE potential   
naturally contains a combination of 
spin-spin and tensor terms.  
One typically neglects the spin-orbit ($LS$) potential 
which appears 
in a non-relativistic reduction of the 
OGE amplitude. 
The pragmatic reason is that the mass spectrum of the excited 
baryons show no strong evidence for a strong $LS$ potential   
between quarks. 
As a possibility 
Isgur and Karl \cite{ik78} argued that 
the $LS$-potential from the effective OGE 
could be cancelled by the $LS$-potential from 
a Lorentz-scalar confining interaction of the quarks. 
This cancellation is shown to occur  
in a chiral (cloudy) bag-model description 
of the negative parity $N^*$ and 
$\Delta^*$ states \cite{mw84}, 
see erratum in Ref.\cite{um89}. 
The 
{\it spin-independent} confining HO potential, 
implies that the two $p$-states with $j=$ 1/2 and 3/2 
are degenerate in energy. 
The OGE spin-dependent potential contributes to the 
$SU(3)$-multiplet mixing in the 
excited baryon states.  
Isgur and Karl observed that most 
physical states are closer to pure {\it uds}-basis 
states (characterized by the two Jacobi coordinates 
and their corresponding HO frequencies) 
than to pure 
$SU(3)$ eigenstates~\cite{ik78}. 

\vspace{3mm}

%%%%%%%%%%%%%%%%%%%%%%%%%%%%%%%%%%%%%%%%%%%%%%%%%%%%%%
%\subsubsection
%{\bf A relativistic quark model }
%%%%%%%%%%%%%%%%%%%%%%%%%%%%%%%%%%%%%%%%%%%%%%%%%%%%%%

\vspace{3mm} 

The Bonn group \cite{bonn06} 
calculates the electromagnetic 
decay of hyperon resonances within a relativistic 
quark model where the propagators of the 
three quarks are given by the two Jacobi momenta. 
In 
their evaluations they propose a 
confinement potential $V_{conf}$ 
with a Dirac structure such that 
$V_{conf}$ reduces to a {\it spin-independent}  
linear confinement potential 
in the non-relativistic limit.  
In addition, Ref. \cite{bonn06} includes 
an instanton induced two-quark potential 
instead of the effective OGE potential. 
These two potentials are included in 
the Bethe-Salpeter equation, which is solved in the 
instantaneous approximations. 
%evaluate $\Gamma_\gamma$. 

\vspace{3mm}

%%%%%%%%%%%%%%%%%%%%%%%%%%%%%%%%%%%%%%%%%%%%%%%%%%%
%\subsection
%{\bf The bag models}
%%%%%%%%%%%%%%%%%%%%%%%%%%%%%%%%%%%%%%%%%%%%%%%%%%%%

In bag-models the (Lorentz scalar) 
confinement condition introduces 
a strong $LS$ splitting of the quark eigenstates.  
The quark $p$-state with $j=3/2$ ($p3/2$) 
has a lower energy 
than the $j=1/2$ ($p1/2$) state which affect 
the two  corresponding 
eigenfunctions. 
Furthermore, 
this effective $LS$ component introduces an $SU(3)$  
multiplet mixing in the 
non-perturbed wave functions of 
the excited baryon in addition to the 
mixing due to the effective OGE discussed above. 
The pion and kaon clouds in 
the chiral (cloudy) bag model mix the  $SU(3)$ 
multiplets as well. 
As mentioned above, in the bag model the 
OGE automatically contains an effective 
$LS$ force of 
opposite sign and about the same magnitude 
compared to the effective 
$LS$ of the bag confinement 
condition~\cite{mw84,um89}. 
In bag models the $u$- and $d$ quarks 
are massless whereas the $s$-quark is massive, 
i.e. the explicit breaking of flavor-$SU(3)$ is 
built into the excited baryon wave functions. 
This is however not the $uds$-basis discussed by  
Isgur and Karl. 
Unlike the NRQM it is highly non-trivial 
to introduce the Jacobi coordinates for  
the three quark wave functions of bag models 
which are described by a product of Dirac spinors. 
In the bag model the $\Lambda (1116)$ translational 
mode 
%(assumed to be the {\bf 56}-plet) 
is projected out 
following the procedure of 
DeGrand et al. \cite{DeGrand76}.  
In this procedure one transforms the three-quark 
$jj$ coupled states to 
the $LS$-basis and 
assumes that the 
$SU(6)\times O(3)$ $L=1, {\bf 56}$-plet is 
the translational mode of the 
hyperon ground state, $\Lambda (1116)$. 
The baryon mass spectrum appears to be well 
described in (cloudy/chiral) 
bag models despite the strong $LS$ 
confinement component, see 
Refs.~\cite{um89,mw84}.

\vspace{3mm} 

%%%%%%%%%%%%%%%%%%%%%%%%%%%%%%%%%%%%%%%%%%%%%%%%%%%%%%%%%%%
%
{\bf The electromagnetic decay rate} 
%%%%%%%%%%%%%%%%%%%%%%%%%%%%%%%%%%%%%%%%%%%%%%%%%%%%%%%%%%

The various theoretical models generate 
very different values for 
$\Gamma_\gamma$, which is very sensitive to  
the amount of  admixture of higher configuration 
states in the hyperon ground state. 
Another important factor, which can change the value of  
$\Gamma_\gamma$, 
is the amount of $SU(3)$- 
octet admixtures in $\Lambda(1520)$, 
i.e. the magnitude of the coefficients 
$b$ and $c$ in Eq.(\ref{eq:L1520}), 
see e.g. the discussion in Ref.~\cite{kms85}. 
In Table 1 the first column indicate 
the quark model used in calculating $\Gamma_\gamma$ 
and the 
second column gives the corresponding reference. 
The values of the 
coefficients in Eq.(\ref{eq:L1520}) used in the 
different model calculations 
are presented in the next three columns. 
The sixth column gives which 
$\Lambda (1116)$ wave function is used, 
Eq.(\ref{eq:Lsymm}) or Eq.(\ref{eq:Lmixed}), and the 
last column gives the value for the 
decay rate evaluated in the different references. 
%
%%%%%%%%%%%%%%%%%%%%%%%%%%%%%%%%%%%%%%%%% 
% 
As is evident from Table 1, most calculations 
adopt the values for $a$, $b$ and $c$ as 
determined by Isgur and Karl~\cite{ik78},
$a=0.91$, $b=0.01$ and $c=0.40$. 
The bag model results for the 
values of the coefficients, $a$, $b$ and $c$, 
differ from the NRQM of Isgur and Karl mainly due 
to the bag confinement condition. 
The three coefficients in the row 
labeled ``MIT bag" in Table 1 are determined from 
the expression in Eq.(4) of Ref.~\cite{kms85} 
using 
the definitions Eqs.(4.6) and (B.4) 
in Ref.\cite{mw84}. 
In the chiral/cloudy bag model evaluations 
the meson cloud contribute to further change
the values of the three 
coefficients.\footnote{ 
The bag model values of the three  
coefficients are extracted 
using the procedure of Ref.~\cite{DeGrand76}: 
After a recoupling from the relativistic 
$jj$-basis to the 
$LS$-basis, the {\bf 56}-plet is 
projected out. 
It is then possible to determine the multiplet 
mixing coefficients, $a$, $b$ and $c$, 
in Eq.(\ref{eq:L1520}) given in Table 1. 
}

\vspace{3mm} 

%%%%%%%%%%%%%%%%%%%%%%%%%%%%%%%%%%%%%%%%%%%%%%%%%%%%%%%
  \begin{center}
  Table 1 : \parbox[t]{5.3in}{
The value of 
$\Gamma_\gamma$ from various publications 
are given in the last column. 
The coefficients $a$, $b$ and $c$  
  of the $\Lambda (1520)$ wave functions, 
Eq.(\ref{eq:L1520}) 
used in the various publications are given. 
The version of the ground state wave function 
$\Lambda (1116)$, 
Eq.(\ref{eq:Lsymm}) or Eq.(\ref{eq:Lmixed}), 
used in the different 
evaluations of $\Gamma_\gamma$ is indicated in the 
next to last column. 
The ``dash" means that values cannot be ascertained. 
See text for details.} 
\end{center}
  $$
  \begin{array}{|l|r | |r|r|r|r|r|}
  \hline
 {\rm Models } & {\rm Ref.}& a & b &
c & \Lambda (1116) & \Gamma_{\gamma} {\rm (keV)} \\
  \hline
  \hline
 {\rm NRQM}\; \; \; &\cite{dhk83} & 
0.91& 0.01& 0.40& {\rm Eq.}(\ref{eq:Lsymm}) & 96
\\ \hline 
{\rm NRQM}\; (SU(6){\rm -basis})\; &\cite{kms85} & 
0.91& 0.01& 0.40& -  & 98 
\\ \hline 
%
%{\rm NRQM} & \cite{copley69}& -& -& -& 
%{\rm Eq.}(\ref{eq:Lsymm}) & 100
%\\ \hline 
%
\chi{\rm QM} \; \; \; &\cite{Yu06} & 
0.91 & 0.01 & -0.40 &{\rm Eq.}(\ref{eq:Lsymm})^* 
&  85
\\ \hline 
\chi{\rm QM}\; \; \; &\cite{Yu06PRD} & 
0.91 & 0.01 & -0.40 &{\rm Eq.}(\ref{eq:Lmixed})^* 
&  134 
\\ \hline 
{\rm NRQM}\; (uds{\rm -basis})\; &\cite{kms85} & 
- & - & - & {\rm Eq.}(\ref{eq:Lmixed}) & 154 
\\ \hline 
{\rm MIT}\; {\rm bag}\; &\cite{kms85} &  
 0.86 & 0.34 & -0.37 %{\rm Eq.}(\ref{eq:Lmixed}) 
& - & 46 
\\ \hline
{\rm Chiral/Cloudy}\; {\rm bag}\; & \cite{um91,um93} &
-0.95  &  -0.09& 0.29 & {\rm Eq.}(\ref{eq:Lsymm}) & 32 
\\ \hline 
{\rm RCQM } \; \; \; &\cite{wpr91} & 
 0.91 &  0.01& 0.40 & {\rm Eq.}(\ref{eq:Lmixed}) & 215 
 \\ \hline 
{\rm Bonn-CQM } \; \; \; &\cite{bonn06} & 
 - &  - & - & - 
& 258 
 \\ \hline 
\end{array}
$$
%%%%%%%%%%%%%%%%%%%%%%%%%%%%%%%%%%%%%%%%%%%%%%%%%%%%%%%

\vspace{3mm}

%%%%%%%%%%%%%%%%%%%%%%%%%%%%%%%%%%%%%%%%%
%\section
{\bf Discussion and Conclusions} 
%%%%%%%%%%%%%%%%%%%%%%%%%%%%%%%%%%%%%%%%%%%%%%%%%%%%%% 

\vspace{3mm}

Table 1 shows how the calculated values of 
$\Gamma_\gamma$  varies according to which 
wave function for $\Lambda (1116)$ is used. 
The standard $SU(6)$-basis evaluations of 
Refs.~\cite{dhk83,kms85,Yu06}, in the 
first three rows are very similar and they find 
$\Gamma_\gamma \approx 100$ keV, confirming  
the early  evaluation of 
$\Gamma_\gamma = 100$ keV by 
Copley {\it et al.}~\cite{copley69}.
It should be noted that 
Refs.~\cite{Yu06,Yu06PRD} ``dress" the  
constituent quark wave function with 
Goldstone bosons 
(denoted by a $*$ in Table 1). 
In these three first rows  
the hyperon ground state wave function 
appears to be given by Eq.(\ref{eq:Lsymm}).  
All the three values of $\Gamma_\gamma$ 
are small compared to the experimental value 
$\sim 160$ keV. 

\vspace{3mm}

The influence of the configuration mixing in  
the hyperon ground state, Eq.(\ref{eq:Lmixed}),  
due to the effective color 
hyperfine interaction \cite{ik78} is evident 
if we compare  the values of 
$\Gamma_\gamma$ in the three first rows with 
the fourth row $\Gamma_\gamma$ from 
Ref.\cite{Yu06PRD}, 
the published 
version of Ref.\cite{Yu06}. 
In the fifth row 
$\Gamma_\gamma $ 
%= 154$ keV 
is evaluated in 
the $uds$ basis by 
Kaxiras {\it et al.}~\cite{kms85} 
where Eq.(\ref{eq:Lmixed}) 
enters naturally.   
In other words, when configuration mixing in both 
$\Lambda (1520)$ and $\Lambda (1116)$ are included, 
the calculated values of $\Gamma_\gamma$ 
appear to be close to the experimental 
value of $\sim 160$ keV. 

\vspace{3mm}

The importance of configuration mixing in 
$\Lambda (1116)$ 
is also 
found by Warns {\it et al.}~\cite{wpr91}.
Refs.~\cite{wpr91} and \cite{bonn06} 
evaluate $\Gamma_\gamma$ 
using a relativistic approach. 
Warns {\it et al.}~\cite{wpr91} 
employ a relativistic constituent 
quark model (RCQM) and quote a (lowest order) 
non-relativistic result  
$\Gamma_\gamma= 48$ keV. 
As a relativistic correction Ref.~\cite{wpr91} 
considers the configuration 
mixing generated by the color hyperfine 
interaction (OGE) of 
Isgur and Karl, which results in a   
configuration mixing of 
$\Lambda (1116)$, Eq.(\ref{eq:Lmixed}). 
Furthermore, the center-of-mass recoil effect  
is also counted as a relativistic correction 
by \cite{wpr91}. 
The relativistic ``full" 
RCQM result of Ref.\cite{wpr91} is shown in Table 1.
As mentioned, 
Van Cauteren {\it et al.}~\cite{bonn06} use the 
Bethe-Salpeter equation to account for 
relativity in evaluating, e.g. $\Gamma_\gamma$.  
The recoil boost of $\Lambda (1116)$ is an 
integral part in this  
evaluation.  
%Ref. \cite{bonn06}. 
Unfortunately, 
these two references \cite{wpr91,bonn06} 
do not indicate the magnitudes 
of the different relativistic corrections.  
It is not clear why these two publications, 
Refs.~\cite{wpr91,bonn06}, find values 
so much larger than the other model evaluations of 
$\Gamma_\gamma$. 

\vspace{3mm} 

The bag models clearly predict too small decay rates. 
One reason for 
this 
%the small $\Gamma_\gamma$ predicted by the bag models 
could be that 
the quark energy differences 
$E_{p3/2} - E_{s1/2} \simeq 230$ MeV and 
$E_{p1/2} - E_{s1/2} \simeq 350$ MeV 
(bag radius $\simeq 1$ fm) 
are 
both smaller than the photon 
energy $E_\gamma \simeq 400$ MeV. 
%of this hyperon decay. 
%
These two energy differences reflect  
the strong spin-orbit splitting of 
the bag confinement condition and are built into 
the quark wave functions. 
%
%(In potential models the quark wave functions 
%do not contain  
%energy difference between the two p-states.)  
% 
%  
The electromagnetic transition of  
a quark from one of the two $p$-states 
to  the ground state, $s1/2$  is 
evaluated using the relativistic 
bag wave functions. 
To project out the translational mode the 
$SU(6)$ argument, which was discussed earlier and found 
to be questionable, was  
implemented.  
In addition, Refs.\cite{um91,um93} used 
Eq.(\ref{eq:Lsymm}) to describe 
$\Lambda (1116)$ and  furthermore 
the boost of the ground state 
$\Lambda (1116)$ (recoil momentum  
$p_\Lambda \simeq 400$ MeV/c) 
was neglected. 
The small difference between the MIT bag and the 
Chiral (Cloudy) bag model results are mainly 
due to two factors. 
In the Chiral/Cloudy bag all OGE 
contributions involving the 
$p3/2$ quark state are included 
and   
an additional small contribution 
to the decay rate \cite{um93} comes from 
the meson cloud 
(pions and kaons). 
This small meson cloud contribution to $\Gamma_\gamma$ 
is confirmed by 
Ref. \cite{oset06},  
which uses a unitary extension 
of chiral perturbation theory to evaluate 
$\Gamma_\gamma$ 
in a meson-baryon coupled channels approach.   

\vspace{3mm}

As presented above the 
electromagnetic decay rates discriminate easily  
between the different model descriptions 
of the hyperons. 
In order to make further progress in 
our understanding of the structure of the 
baryon states, 
it is necessary to experimentally determine several  
other radiative decay rates from  
excited baryon states. 
There also exists 
one serious theoretical problem for quark models, 
namely 
their prediction that 
the lightest $J^P=1/2^-$-state, $\Lambda (1405) $ 
and the lightest $J^P=3/2^-$-state, $\Lambda (1520) $  
are almost degenerate states, 
see e.g. \cite{ik78,close79,ki80,um89,um91,bonn06}.  
This might indicate that $\Lambda (1405)$ 
is a special state containing a dominant 
$\bar{q}q^4$ (or meson-baryon) component. 
As originally proposed~\cite{Dalitz}, a full 
coupled channel treatment 
of a meson-baryon system coupled 
to the ``bare" $q^3$ states with the quantum 
numbers of both 
$\Lambda (1405)$ and $\Lambda (1520)$ could possibly 
resolve this (almost) mass-degeneracy found 
in quark model calculations. 

\vspace{3mm}

To summarize the 
$uds$-basis for 
the NRQM as advocated by   
Isgur and Karl \cite{ik78} supplemented 
by an effective color hyperfine interaction, 
predicts an important 
configuration mixing of 
$\Lambda (1116)$ in addition to the mixing of 
the flavor components of $\Lambda (1520)$. 
It appears that these non-trivial 
mixings are needed to explain the 
observed large value of  
$\Gamma_\gamma$.  
The evaluations of $\Gamma_\gamma$ in 
Refs.~\cite{kms85} and \cite{Yu06PRD} 
include the mixed states of Isgur and Karl in both the 
initial and final hyperons and their values for 
$\Gamma_\gamma$ are consistent with the 
experimental value, see 
Refs.~\cite{CLAS05,russian04,mast68}.  
Ref.~\cite{wpr91}  
also includes the 
configuration mixing in both hyperons, 
and find a 
result that lies within the experimental 
errors of the CLAS collaboration result\cite{CLAS05}. 

\vspace{3mm} 

One conclusion to be drawn is that 
flavor $SU(3)$ breaking as well as higher quark states 
configuration mixing of the ground state baryons    
have to be an 
integral part of the 
wave function of the baryons, 
i.e. we should use the $uds$-basis to evaluate  
the three-quark wave functions of the baryons. 
The $SU(3)$ breaking is most easily ``visualized" 
in the excited hyperon states: 
given that the $s$-quark is heavy compared to the 
$u$- and $d$-quarks (this mass difference is 
dramatic in bag models), a scenario emerges 
where the pair of light quarks ``orbits" 
the heavy $s$-quark in the excited hyperons. 
Another tentative conclusion, which is 
based on the bag model results, 
is that the 
confinement mechanism 
should not have a strong spin-orbit component. 
Furthermore, in 
some models the procedure for removing the 
center-of-mass coordinate has to be re-examined, 
as discussed in, e.g. Ref.\cite{um89}. 
Flavor $SU(3)$ is broken and therefore  
the $SU(6)$ {\bf 56}-multiplet 
is not representative of the translational mode  of 
the baryon ground state. 

\vspace{3mm}

{\bf Acknowledgements}

The author is grateful for valuable 
comments from R. Gothe and K. Kubodera. 
This work is supported in part by an NSF grant.

%
%%%%%%%%%%%%%%%%%%%%%%%%%%%%%%%%%%%%%%%%%%%%%%%%%%%%%%%%%%
%\newpage 
%%%%%%%%%%%%%%%%%%%%%%%%%%%%%%%%%%%%%%%%%%%%%%%%%%%%%%%%%%

\end{document}